\batchmode
\makeatletter
\def\input@path{{/Users/axelaraneda/Desktop/Research/Portfolio/}}
\makeatother
\documentclass[english]{article}
\usepackage{lmodern}
\usepackage[T1]{fontenc}
\usepackage[latin9]{inputenc}
\usepackage[a4paper]{geometry}
\usepackage{babel}
\usepackage{booktabs}
\usepackage{units}
\usepackage{url}
\usepackage{amsmath}
\usepackage{amsthm}
\usepackage{amssymb}
\usepackage{graphicx}
\usepackage[authoryear,round]{natbib}
\usepackage[pdftex,unicode=true]
 {hyperref}

\makeatletter

\providecommand{\tabularnewline}{\\}

\newcommand{\lyxaddress}[1]{
	\par {\raggedright #1
	\vspace{1.4em}
	\noindent\par}
}

\usepackage{hyperref}
\date{}
\usepackage[format=hang,font=small,labelfont=bf]{caption}

\@ifundefined{showcaptionsetup}{}{%
 \PassOptionsToPackage{caption=false}{subfig}}
\usepackage{subfig}
\makeatother

\begin{document}
\title{\textbf{Asset volatility forecasting:}\linebreak{}
\textbf{The optimal decay parameter in the EWMA model}\thanks{This paper has benefited from the comments of the participants of
both the ``11\textsuperscript{th} Biennial Conference of the Czech
Economic Society'' and the joint conference ``European Working Group
for Commodities and Financial Modelling 63\textsuperscript{rd} Meeting
\& XVIII International Conference on Finance and Banking FI BA 2021''.}}
\author{Axel A.~Araneda\thanks{Email: \protect\href{mailto:axelaraneda@mail.muni.cz}{\texttt{axelaraneda@mail.muni.cz}}} }
\maketitle

\lyxaddress{\begin{center}
\vspace{-2.5em}Institute of Financial Complex Systems\\Faculty
of Economics and Administration\\ Masaryk University\\Lipov\'a
41a, 602 00 Brno, Czech Republic.
\par\end{center}}

\begin{center}
\vspace{-2em} 
\par\end{center}
\begin{abstract}
The exponentially weighted moving average (EMWA) could be labeled
as a competitive volatility estimator, where its main strength relies
on computation simplicity, especially in a multi-asset scenario, due
to dependency only on the decay parameter, $\lambda$. But, what is
the best election for $\lambda$ in the EMWA volatility model? Through
a large time-series data set of historical returns of the top US large-cap
companies; we test empirically the forecasting performance of the
EWMA approach, under different time horizons and varying the decay
parameter. Using a rolling window scheme, the out-of-sample performance
of the variance-covariance matrix is computed following two approaches.
First, if we look for a fixed decay parameter for the full sample,
the results are in agreement with the RiskMetrics suggestion for 1-month
forecasting. In addition, we provide the full-sample optimal decay
parameter for the weekly and bi-weekly forecasting horizon cases,
confirming two facts: i) the optimal value is as a function of the
forecasting horizon, and ii) for lower forecasting horizons the short-term
memory gains importance. In a second way, we also evaluate the forecasting
performance of EWMA, but this time using the optimal time-varying
decay parameter which minimizes the in-sample variance-covariance
estimator, arriving at better accuracy than the use of a fixed-full-sample
optimal parameter. 



\vspace{2em}
\end{abstract}

\section{Introduction}

Volatility is a key parameter in financial problems as derivative
pricing, portfolio allocation, and value-at-risk. Thus, optimal (and
sometimes efficient) forecasting plays a fundamental role in \emph{ex-ante}
valuation. Despite there are several sophisticated methods to deal
with the volatility estimation, as the ARCH-GARCH family \citep{Bollerslev2010}
or stochastic volatility (SV) models \citep{shephard2009stochastic};
these models are complex and computationally expensive, especially
in a multi-asset framework. Then, and depending on their purposes,
some practitioners prefer to use costless implementations. One of
those is the exponentially weighted moving average (EWMA) model, capable
to address two well-known \textquotedblleft stylized facts\textquotedblright{}
as heteroskedasticity, and volatility clustering. This model assigns
different weights to the past information, where more recent lags
receive more importance than old observations. The weight assignation
decays exponentially at a rate $\lambda$. 

One of the greater advantages of the EWMA over the standard GARCH
models is related to computation simplicity. While in the former,
the variance is easily and quickly updated; in the latter, a new likelihood
maximization should be performed for every run. This point is particularly
problematic in the multivariate context, where MGARCH models become
very time-costly, especially when the number of assets rises. Also,
EWMA captures the autocorrelation of the squared returns better than
standard GARCH approaches \citep{bee2010dynamic}. In terms of drawbacks,
the EWMA model doesn't consider a long-run variance nor mean-reversion;
so it could be not suitable for long-term forecasting. Moreover, the
decay parameter should be adjusted in function of the time horizon.

\citet{ding2010forecasting} point that EWMA exhibits a greater forecasting
accuracy to real data compared to GARCH and SV models. \citet{gonzalez2004forecasting}
found EWMA brings good results in the option pricing context. In addition,
and for portfolio purposes, \citet{zakamulin2015test} reveals that
EWMA performance is very similar to the DCC-GARCH, in both covariance-matrix
forecasting error and portfolio tracking error. 

The EWMA method was popularized by \citet{RM1996}, who recommends
a smoothing factor equal to 0.97 for a 1-month forecast. These values
are obtained minimizing the average squared errors for a large number
of time-series (detailed procedure on section 5.3.2 of the Technical
Document). However, \citet{bollen2015should} and \citet{gonzalez2007optimality}
tested empirically the optimal $\lambda$ under several criteria,
found that the \citeauthor{RM1996} suggested values are overestimated
and far from optimal.

The goal of this paper is to deal with a common issue that should
be faced by the practitioners in their implementation: what should
be the optimal value for the decay parameter in the EMWA model? This
question is answered taking into account different stock returns time-series
and forecasting horizons: weekly, bi-weekly and monthly. For this,
we use large historical data (daily adjusted-closing prices for the
27 years period 1994-2018) of the top blue-chip companies in the US
stock market (DJIA components), and we evaluate the volatility estimation
in a multi-asset context (variance-covariance matrix) of the EWMA
approach under different specifications, measuring its out-of-sample
performance, in a rolling-window scheme. The use of daily data obeys
two reasons. First, for forecasting horizons greater or equal than
a week, as the selected ones, the daily frequency offers similar results
to high-frequency for volatility prediction \citep{lyocsa2021fx}.
And second, the high-frequency data is less accessible, and computationally
(and economically) costly, which is in the opposite direction to this
work aims.  We address and compare two different ways to obtain the
optimal smoothing parameter: i) the use of a full-sample optimal $\lambda$
(RiskMetrics methodology), and ii) the optimal in-sample time-varying
$\lambda$, which maximizes the accuracy of the covariance matrix
previous to the forecasting. We found a better predictive accuracy
for the second way for the selected forecasting horizons

The outline of this document is the following. First, a brief review
of the realized volatility and covariances, and their estimation by
the EMWA method, is developed. After that, in section \ref{sec:Methodology},
the methodology of the paper (data, procedure, and error measure),
is described. The results are displayed and analyzed in section \ref{sec:Results};
and finally, the main conclusions are summarized.

\section{Definitions}

\subsection{Realized Volatility and Covariances}

First, the continuously compounded return (log-return) of the $i-$asset
is defined as:\\
\[
r_{i,t}=\ln\left(\frac{P_{i,t}}{P_{i,t-1}}\right)
\]
\\
\noindent where $P_{i,t}$ is the price of the asset $i$ at time
$t$.

An unbiased estimator for the ex-post true volatility is the realized
volatility. At the day $t$, the realized volatility the $T$ trading
days period  (from $t-T+1$ to $T$), is computed as the square root
of the $T-$ most recent squared daily returns \citep{andersen1999forecasting,bollen2015should}:\\

\begin{eqnarray*}
\sigma_{i,t_{T}} & = & \sigma_{i,\left[t-T+1,t\right]}\\
 & = & \sqrt{\sum_{k=0}^{T-1}r_{i,t-k}^{2}}
\end{eqnarray*}
\\

Later, the entries of the covariance matrix $\boldsymbol{C}_{t_{T}}$
depends on both the volatilities and correlations for the recent $T-$days:\\

\begin{eqnarray*}
\left(\boldsymbol{C}_{t_{T}}\right)_{ij} & = & \sigma_{ij,t_{T}}\\
 & = & \sigma_{i,t_{T}}\sigma_{j,t_{T}}\rho_{ij,t_{T}}
\end{eqnarray*}
\\

\noindent being $\rho_{ij,t_{T}}$ the correlation coefficient among
the returns $i$ and $j$, between the days $t-T+1$ and $t$. Equivalently,
we have:\\

\begin{eqnarray*}
\sigma_{ij,t_{T}} & = & \sum_{k=0}^{T-1}r_{i,t-k}r_{j,t-k}
\end{eqnarray*}
\\

\subsection{EWMA volatility}

The equally weighted moving average model, estimates the one-day-ahead
daily conditional variance as:

.

\begin{equation}
\hat{\sigma}_{i,t+1}^{2}=\left(1-\lambda\right)\sum_{n=0}^{\infty}\lambda^{n}r_{i,t-n}^{2}\label{eq:EWMA}
\end{equation}
\\

\noindent with $0<\lambda<1$. 

If $\lambda$ moves away from 1, the EWMA assigns higher weights to
the recent than the past observations. Then, the quality of the results
depends on the election of the parameter $\lambda$. A value greater
(lower) than the optimal goes to an under-reaction (over-reaction)
to the new information input.

Equivalently, Eq. (\ref{eq:EWMA}) could be expressed by the following
recurrence relation:\\

\begin{equation}
\hat{\sigma}_{i,t+1}^{2}=\left(1-\lambda\right)r_{i,t}^{2}+\lambda\sigma_{i,t}^{2}\label{eq:EWMA2}
\end{equation}
\\

The first term of the RHS of (\ref{eq:EWMA2}) updates the variance
due to the new information, while the second one represents the persistence
effect. It's easily to show that new iterations in \ref{eq:EWMA2},
conditional to the information up to time $t$, won't update the variance
result. Thus, for any $k\in\mathbb{N^{+}}$:\\

\begin{equation}
\hat{\sigma}_{i,\left.t+k+1\right|t}^{2}=\hat{\sigma}_{i,t+1}^{2}\label{eq:future}
\end{equation}
\\

Then, Eq. \ref{eq:future} implies that the volatility estimation
should be scaled to the length of the time horizon; i.e., for the
$T$-day period:\\

\[
\hat{\sigma}_{i,t_{T}}^{2}=T\cdot\hat{\sigma}_{i,t+1}^{2}
\]
\\

Eq. \ref{eq:EWMA2} could be considered as a particular case of the
of \citet{engle1986modelling} IGARCH(1,1), which at the same time
is a restricted case of the \citet{bollerslev1986generalized} standard
GARCH(1,1) such that the coefficients of both lagged squared-returns
and lagged variances sum one (i.e., a unit-root GARCH). This restrictions
drive to an infinite unconditional variance and persistence in the
variance shocks. 

A big feature of the EWMA is its straightforward extension to the
multivariate case, where the covariance forecasting among the assets
$i$ and $j$ is given by:\\

\begin{eqnarray}
\sigma_{ij,t+1} & = & \left(1-\lambda\right)\sum_{n=0}^{\infty}\lambda^{n}r_{i,t-n}r_{j,t-n}\label{eq:Covar}\\
 & = & \left(1-\lambda\right)r_{i,t-n}r_{j,t-n}+\lambda\sigma_{ij,t-n}\label{eq:COVAR_EWMA}
\end{eqnarray}
\\

It should be marked that the simplicity of Eq. (\ref{eq:COVAR_EWMA})
is due to the use of the same $\lambda$ for all the time-series.
It fact, it also guarantees a positive semi-definite variance-covariance
matrix . Besides, the multivariate EWMA could be read as a special
case of the \citet{engle1995multivariate} multivariate GARCH.

We can note from eq. (\ref{eq:EWMA}) that the EWMA use all the past
information (summation up to infinity). However, since the importance
of the past values decays exponentially tending to zero, we can select
an effective number of historical data (cut-off) according to a desired
level of tolerance \citet{RM1996}:\\

\begin{equation}
N=\frac{\ln\left(\Upsilon_{L}\right)}{\ln\left(\lambda\right)}\label{eq:tol}
\end{equation}
\\

\noindent where $N$ is the cutoff point and $\Upsilon_{L}$ the
tolerance level. Thus, Eq. \ref{eq:tol} means that for a given $\lambda$
the information not considered (above the $N$-lag) would have contributed
with a total weight of $\Upsilon_{L}$. 

If we truncate the summation up to $N$ terms, the weights don't sum
the unity, Eqs. (\ref{eq:Covar}) and (\ref{eq:COVAR_EWMA}) should
be corrected as:\\

\begin{equation}
\hat{\sigma}_{ij,t+1}=\frac{\left(1-\lambda\right)}{\left(1-\lambda^{N}\right)}\sum_{n=0}^{N-1}\lambda^{n}r_{i,t-n}r_{j,t-n}\label{eq:truncated}
\end{equation}

\begin{equation}
\hat{\sigma}_{ij,t+1}=\frac{\left(1-\lambda\right)r_{i,t}r_{j,t}+\lambda\sigma_{ij,t}}{\left(1-\lambda^{N}\right)}\label{eq:truncated-1}
\end{equation}
\\

Note that in the limit $\lambda\rightarrow$1, the above equations
converge to the raw-sample variance and covariances using $N$ lags. 

\section{Methodology\label{sec:Methodology}}

\subsection{Data}

We use the continuously compounded returns of the daily adjusted-closing
prices of 22\footnote{PFE, RTX and XOM were excluded from the index in the middle of 2020,
but we also considered them in the analysis.} stocks listed continuously in the Dow Jones Industrial Average (DJIA)
index from 2000 up to 2020 (see Table \ref{tab:Components}). We consider
the historical records from January 3, 1994 to December 31, 2020 (6799
values for each asset) and we examine the forecasting for each day
in the period 2000-2020. This time-span arises crisis, bull markets,
and standard periods. The database is freely available at Yahoo! Finance
(\url{www.finance.yahoo.com}). The data prior to 2000 will be used
for calibration purposes and for exploratory analysis.

\begin{table}
\caption{\textbf{List of the used stocks\label{tab:Components}}}
\vspace{-.5em}
\centering{}%
\noindent\begin{minipage}[t]{1\columnwidth}%
\begin{center}
\begin{tabular}{ccccc}
\cmidrule{1-2} \cmidrule{2-2} \cmidrule{4-5} \cmidrule{5-5} 
\textbf{\footnotesize{}Company} &
 &
 &
\textbf{\footnotesize{}Company} &
\tabularnewline
\textbf{\footnotesize{}Name} &
\textbf{\footnotesize{}Symbol} &
 &
\textbf{\footnotesize{}Name} &
\textbf{\footnotesize{}Symbol}\tabularnewline
\cmidrule{1-2} \cmidrule{2-2} \cmidrule{4-5} \cmidrule{5-5} 
\addlinespace
{\footnotesize{}3M} &
{\footnotesize{}MMM} &
 &
{\footnotesize{}Alcoa Inc.} &
{\footnotesize{}AA}\tabularnewline
{\footnotesize{}American Express} &
{\footnotesize{}AXP} &
 &
{\footnotesize{}Boeing} &
{\footnotesize{}BA}\tabularnewline
{\footnotesize{}Caterpillar Inc.} &
{\footnotesize{}CAT} &
 &
{\footnotesize{}DowDuPont}\footnote{{\footnotesize{}Formerly E.I. Du Pont de Nemours \& Co.}} &
{\footnotesize{}DD}\tabularnewline
{\footnotesize{}ExxonMobil} &
{\footnotesize{}XOM} &
 &
{\footnotesize{}IBM} &
{\footnotesize{}IBM}\tabularnewline
{\footnotesize{}Intel} &
{\footnotesize{}INTC} &
 &
{\footnotesize{}Johnson \& Johnson} &
{\footnotesize{}JNJ}\tabularnewline
{\footnotesize{}JP Morgan Chase} &
{\footnotesize{}JPM} &
 &
{\footnotesize{}McDonald's} &
{\footnotesize{}MCD}\tabularnewline
{\footnotesize{}Merck \& Co.} &
{\footnotesize{}MRK} &
 &
{\footnotesize{}Microsoft} &
{\footnotesize{}MSFT}\tabularnewline
{\footnotesize{}Pfizer} &
{\footnotesize{}PFE} &
 &
{\footnotesize{}Procter \& Gamble} &
{\footnotesize{}PG}\tabularnewline
{\footnotesize{}The Coca-Cola Co.} &
{\footnotesize{}KO} &
 &
{\footnotesize{}The Home Depot} &
{\footnotesize{}HD}\tabularnewline
{\footnotesize{}The Walt Disney Co.} &
{\footnotesize{}DIS} &
 &
{\footnotesize{}Raytheon Tech. Corp.}\footnote{{\footnotesize{}Formerly United Tech. Corp.}} &
{\footnotesize{}RTX}\tabularnewline
{\footnotesize{}Verizon} &
{\footnotesize{}VZ} &
 &
{\footnotesize{}Walmart} &
{\footnotesize{}WMT}\tabularnewline
\cmidrule{1-2} \cmidrule{2-2} \cmidrule{4-5} \cmidrule{5-5} 
\end{tabular}
\par\end{center}%
\end{minipage}
\end{table}

\subsection{Exploratory analysis (1994-1999)}

Table \ref{tab:Descriptive-statistics-1} shows some common statistical
tests over the log-return series for each one of the considered assets,
inside the six-year period 1994-1999. The W-values of the Shapiro-Wilk
test exhibits non-normality in the return time-series. In terms of
ARCH disturbances, Engle's Lagrange multiplier test considering 20
lags, indicates the presence of heteroskedasticity. Besides, the Augmented
Dickey-Fuller (ADF) unit-root test reveals stationarity for all the
variables.

\begin{table}
\textbf{\caption{\textbf{Statistics for daily returns from 2Jan1994 to 31dec1999.\label{tab:Descriptive-statistics-1}}}
}
\centering{}{\footnotesize{}}%
\noindent\begin{minipage}[t]{1\columnwidth}%
\begin{center}
{\footnotesize{}}%
\begin{tabular}{ccccccccc}
\cmidrule{1-4} \cmidrule{2-4} \cmidrule{3-4} \cmidrule{4-4} \cmidrule{6-9} \cmidrule{7-9} \cmidrule{8-9} \cmidrule{9-9} 
\textbf{\footnotesize{}Asset} &
\textbf{\footnotesize{}SW} &
\textbf{\footnotesize{}ARCH(20)} &
\textbf{\footnotesize{}ADF} &
 &
\textbf{\footnotesize{}Asset} &
\textbf{\footnotesize{}SW} &
\textbf{\footnotesize{}ARCH(20)} &
\textbf{\footnotesize{}ADF}\tabularnewline
\cmidrule{1-4} \cmidrule{2-4} \cmidrule{3-4} \cmidrule{4-4} \cmidrule{6-9} \cmidrule{7-9} \cmidrule{8-9} \cmidrule{9-9} 
\addlinespace
{\footnotesize{}AA} &
{\footnotesize{}0.97{*}} &
{\footnotesize{}123.3{*}} &
{\footnotesize{}-34.1{*}} &
 &
{\footnotesize{}AXP} &
{\footnotesize{}0.97{*}} &
{\footnotesize{}311.37{*}} &
{\footnotesize{}-35.1{*}}\tabularnewline
{\footnotesize{}BA} &
{\footnotesize{}0.91{*}} &
{\footnotesize{}46.4{*}{*}} &
{\footnotesize{}-35.6{*}} &
 &
{\footnotesize{}CAT} &
{\footnotesize{}0.97{*}} &
{\footnotesize{}46.1{*}{*}} &
{\footnotesize{}-34.9{*}}\tabularnewline
{\footnotesize{}DD} &
{\footnotesize{}0.98{*}} &
{\footnotesize{}138.6{*}} &
{\footnotesize{}-32.3{*}} &
 &
{\footnotesize{}DIS} &
{\footnotesize{}0.97{*}} &
{\footnotesize{}167.4{*}} &
{\footnotesize{}-37.1{*}}\tabularnewline
{\footnotesize{}HD} &
{\footnotesize{}0.98{*}} &
{\footnotesize{}143.4{*}} &
{\footnotesize{}-36.7{*}} &
 &
{\footnotesize{}IBM} &
{\footnotesize{}0.94{*}} &
{\footnotesize{}13.1}\footnote{{\footnotesize{}No ARCH effects are found using 20 lags. However,
if considering only one lag, the null hypothesis is rejected at the
10\% significance level. This fact is confirmed by the Ljung--Box
$Q$ test to detect autocorrelation over the squared returns; where
$Q^{2}(20)$ provides p<0.0001.}} &
{\footnotesize{}-35.6{*}}\tabularnewline
{\footnotesize{}INTC} &
{\footnotesize{}0.99{*}} &
{\footnotesize{}33.7{*}{*}{*}} &
{\footnotesize{}-37.1{*}} &
 &
{\footnotesize{}JNJ} &
{\footnotesize{}0.99{*}} &
{\footnotesize{}65.0{*}} &
{\footnotesize{}-34.5{*}}\tabularnewline
{\footnotesize{}JPM} &
{\footnotesize{}0.97{*}} &
{\footnotesize{}202.8{*}} &
{\footnotesize{}-32.6{*}} &
 &
{\footnotesize{}KO} &
{\footnotesize{}0.97{*}} &
{\footnotesize{}136.0{*}} &
{\footnotesize{}-34.6{*}}\tabularnewline
{\footnotesize{}MCD} &
{\footnotesize{}0.97{*}} &
{\footnotesize{}135.3{*}} &
{\footnotesize{}-35.1{*}} &
 &
{\footnotesize{}MMM} &
{\footnotesize{}0.97{*}} &
{\footnotesize{}56.1{*}} &
{\footnotesize{}-35.5{*}}\tabularnewline
{\footnotesize{}MRK} &
{\footnotesize{}0.98{*}} &
{\footnotesize{}39.1{*}{*}{*}} &
{\footnotesize{}-36.2{*}} &
 &
{\footnotesize{}MSFT} &
{\footnotesize{}0.99{*}} &
{\footnotesize{}65.0{*}} &
{\footnotesize{}-38.1{*}}\tabularnewline
{\footnotesize{}PFE} &
{\footnotesize{}0.99{*}} &
{\footnotesize{}114.0{*}} &
{\footnotesize{}-33.8{*}} &
 &
{\footnotesize{}PG} &
{\footnotesize{}0.99{*}} &
{\footnotesize{}181.1{*}} &
{\footnotesize{}-36.6{*}}\tabularnewline
{\footnotesize{}RTX} &
{\footnotesize{}0.98{*}} &
{\footnotesize{}162.9{*}} &
{\footnotesize{}-32.7{*}} &
 &
{\footnotesize{}VZ} &
{\footnotesize{}0.98{*}} &
{\footnotesize{}126.5{*}} &
{\footnotesize{}-40.1{*}}\tabularnewline
{\footnotesize{}VMT} &
{\footnotesize{}0.99{*}} &
{\footnotesize{}91.8{*}} &
{\footnotesize{}-37.0{*}} &
 &
{\footnotesize{}XOM} &
{\footnotesize{}0.99{*}} &
{\footnotesize{}102.4{*}} &
{\footnotesize{}-36.3{*}}\tabularnewline
\cmidrule{1-4} \cmidrule{2-4} \cmidrule{3-4} \cmidrule{4-4} \cmidrule{6-9} \cmidrule{7-9} \cmidrule{8-9} \cmidrule{9-9} 
\addlinespace[0.5em]
\end{tabular}{\footnotesize\par}
\par\end{center}
\begin{center}
{\footnotesize{}\vspace{-1.75em} {*} Significant at $p$<0.0001,
{*}{*} Significant at $p$<0.005, {*}{*}{*} Significant at $p$<0.05}{\footnotesize\par}
\par\end{center}%
\end{minipage}{\footnotesize\par}
\end{table}

\subsection{Procedure }

First, we select as forecast horizons: 1week (5 days), 2 weeks (10
days), and 1 month (21 days). Besides, we consider 99 different values
for $\lambda$, from 0.01 to 0.99. We predict the 21 years period
from 2000-2020.

Then, for a given $\lambda$, the first $T$-covariance-matrix estimator
will be given (through the EWMA approach) for the first weekday of
2000 year (Jan 3) using the information available $T$ days back.
The forecasted values will be compared with the $T$-realized-covariances
on that day. Later, the procedure continue using a rolling-window
framework up to the last considered day (December 31, 2020). Thus,
we forecast 253 covariances, for each value of $\lambda$, and for
each working day in the 21 years interval 2000-2020 ($\sim1.32\times10^{8}$
computations).

\subsection{Error measure}

As pointed by \citet{patton2011volatility}, one of the robust error
measures consistent with a noise volatility proxy \nobreakdash-as
the sum of squared returns\nobreakdash- is the mean squared error
(MSE). Since the original \citeauthor{RM1996} methodology uses the
MSE as the main source, the results will be presented considering
that loss function.

Let $\hat{\boldsymbol{C}}_{t_{T}}\left(\lambda\right)=[\hat{\sigma}_{ij,t_{T}}\left(\lambda\right)]$
the $T-$period forecasted variance-covariance matrix at time $t$
for a given $\lambda$, and $\boldsymbol{C}_{t_{T}}=[\sigma_{ij,t_{T}}]$
the realized one. The square forecasted error for the period $T_{k}$
is given by\footnote{Considering only the upper triangular sections of $\hat{\boldsymbol{C}}_{t_{T}}$
and $\boldsymbol{C}_{t_{T}}$, the double accounting of the errors
is avoided.} \citep{zakamulin2015test}:\\

\[
SE_{t_{T}}\left(\lambda\right)=\sum_{i=1}^{\#A}\sum_{j=1}^{i}\left(\hat{\sigma}_{ij,t_{T}}\left(\lambda\right)-\sigma_{ij,t_{T}}\right)^{2}
\]
\\

\noindent where \#$A$ =22 is the number of assets considered. Then,
the MSE is computed averaging among all the periods (windows):\\

\[
MSE_{T}\left(\lambda\right)=\frac{1}{\#W}\sum_{k=1}^{\#W}SE_{t_{T}}\left(\lambda\right)
\]
\\

\noindent being $\#W$ the number of rolling windows (5284).

The optimal $\lambda$ at time $t$, is who provides the minimum squared
error in each date:\\

\[
\lambda_{t_{T}}^{*}=\arg\min\left[SE_{T}\left(\lambda\right)\right]
\]
\\

On the other hand, the optimal smoothing parameter for the full-sample
minimizes the $MSE_{T}$:\\

\[
\lambda_{T}^{*}=\arg\min\left[MSE_{T}\left(\lambda\right)\right]
\]
\\

\subsection{Predictive accuracy}

In order to make a pair-wise comparison of two sets of predicted values,
we employ the widely-used \citet{diebold1995comparing} test who reveals
statistically significant differences in the forecasting accuracy. 

A brief description of the test is addressed in the following. Let
$d_{t}=SE_{a_{t}}-SE_{b_{t}}$ the loss differential, at time $t=\left\{ 1,2,\ldots,N\right\} $,
between the squared errors of the approaches $a$ and \textbf{$b$}.
The Diebold-Mariano (DB) test identify if the predicted values have
the same accuracy or not. Then, the null hypothesis correspond to
zero expected value in the loss differential for all $t$; i.e., $H_{0}$:
$\mathbb{E}\left(d_{t}\right)=0$. In consequence, the alternative
hypothesis considers different levels of accuracy in the forecasting.
Defining $D$ and $\bar{d}$ as the auto-covariance and sample mean
of $d_{t}$, respectively; the t-statistics is given by $DM=\nicefrac{\bar{d}}{\sqrt{\nicefrac{D}{N}}}$.
This test assumes normal distribution under the null hypothesis, $DM\sim N(0,1)$.
So, the null hypothesis is rejected if $\left|DM\right|>z_{\alpha/2}$
being $z_{\alpha/2}$ the upper $z$-value of the standard normal
distribution for the half of the desired level of tolerance $\alpha$.

\section{Results\label{sec:Results}}

We have computed the $T$-variance-covariance estimator by EWMA, varying
the decay parameter $\lambda=\left\{ 0.01,0.02,\ldots,0.99\right\} $
and forecasting horizon $T=\left\{ 5,10,21\right\} $, for each business
day from Jan/03/2000 to Dec/31/2000 using the data available up to
$T$ days earlier.

\begin{figure}
\caption{\textbf{Squared loss function for monthly forecasting\label{fig:Squared-week-1}}}
\subfloat[\textbf{Optimal $\lambda$ over time\label{fig:Optimal-weekly-1}}]%
{\includegraphics[width=0.5\textwidth]{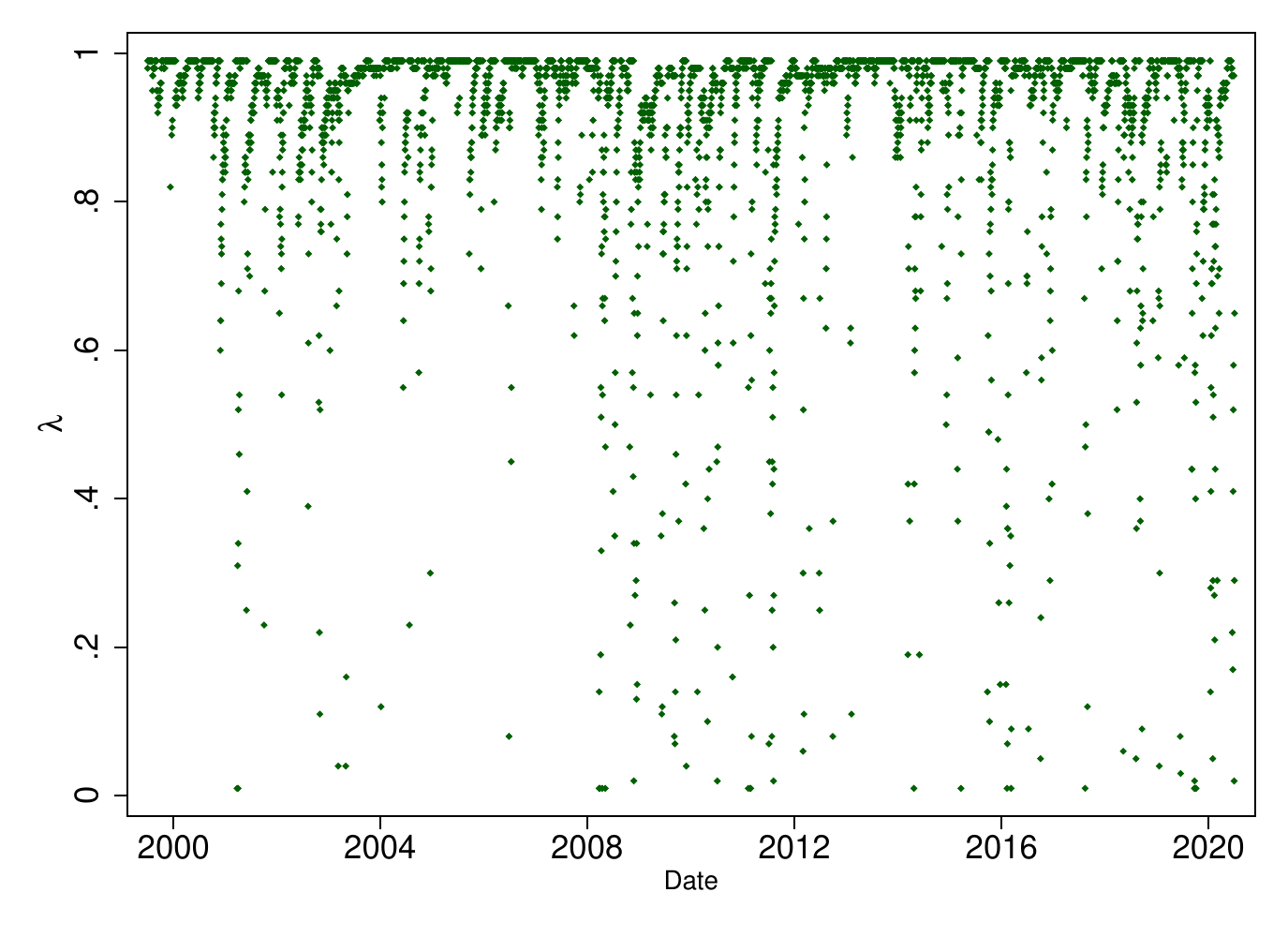}

}%
\subfloat[\textbf{MSE using a fix $\lambda$\label{fig:Mean-squared-errors-Weekly-1}}]%
{\includegraphics[width=0.5\textwidth]{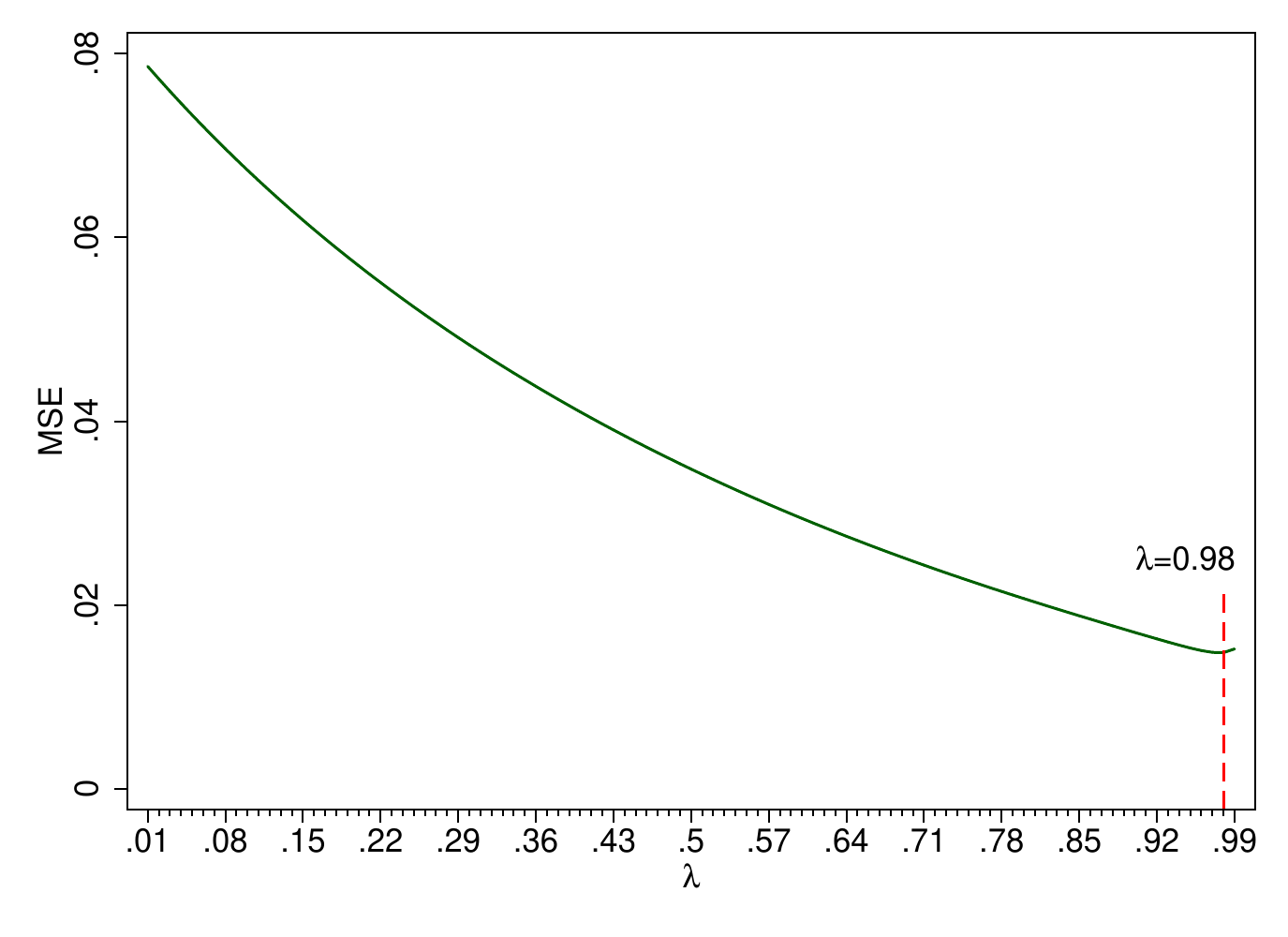}}
\end{figure}

For the 1-month forecasting horizon, we found an optimal full-sample
decay parameter equals 0.98 (see Fig. \ref{fig:Mean-squared-errors-Weekly-1})
which corresponds to a MSE equal to 0.01486. The recommended value
given by \citeauthor{RM1996}, equal to 0.97, ranks in second place
with a very little increase in the MSE value (0.01489). As could be
anticipated by this minimal difference, the \citeauthor{diebold1995comparing}
test couldn't reject the null hypothesis of equal accuracy ($p$-value
>0.9) for the EWMA model using these two smoothing parameters. 

\begin{figure}
\caption{\textbf{MSE for one-week and two-week forecasting using a fix $\lambda$\label{fig:Squared-week-2}}}
\subfloat[\textbf{One-week\label{fig:One-week-forecast}}]%
{\begin{centering}
\includegraphics[width=0.5\textwidth]{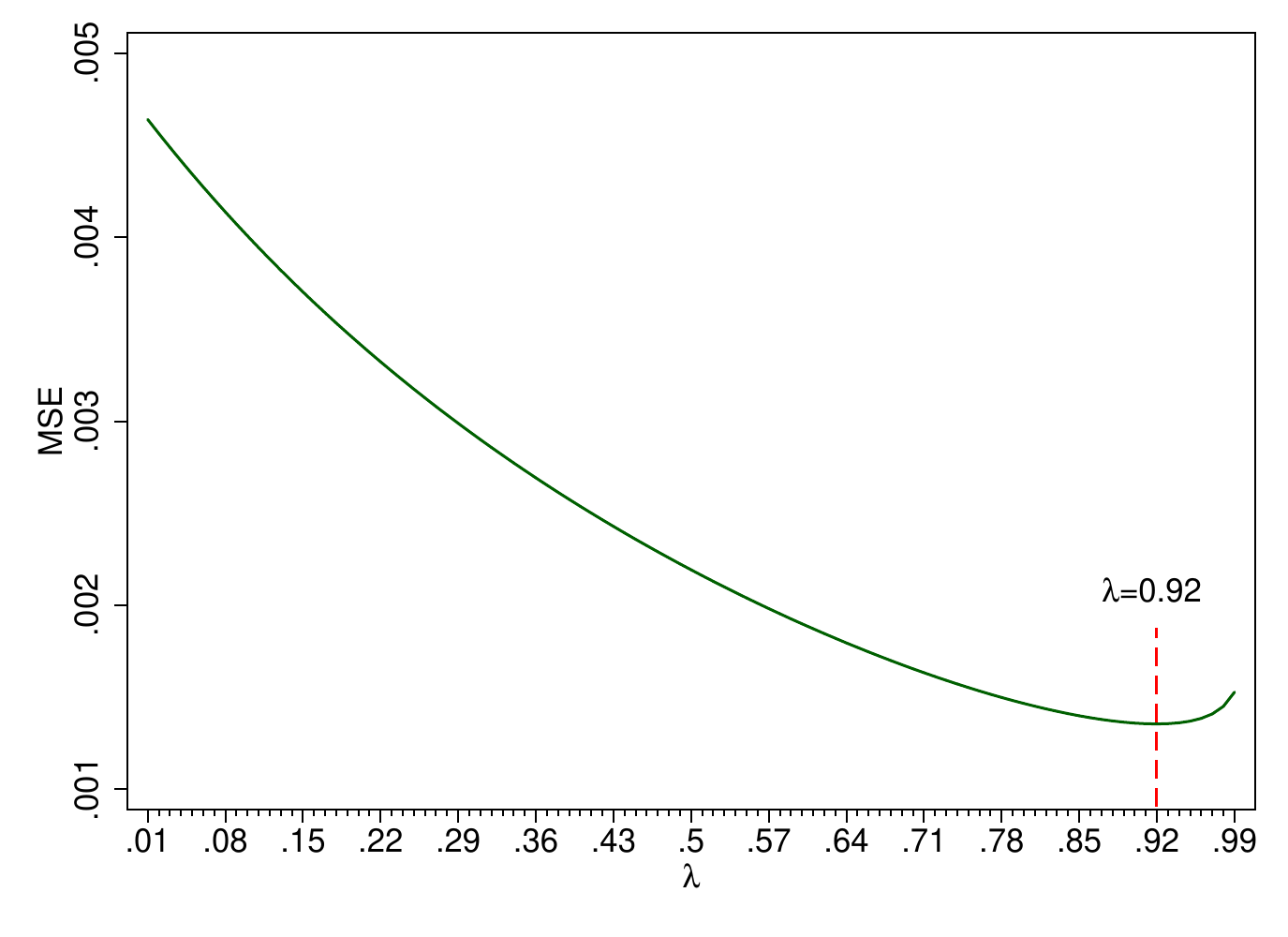}
\par\end{centering}
}%
\subfloat[\textbf{Two-week \label{fig:Two-week-forecast}}]%
{\includegraphics[width=0.5\textwidth]{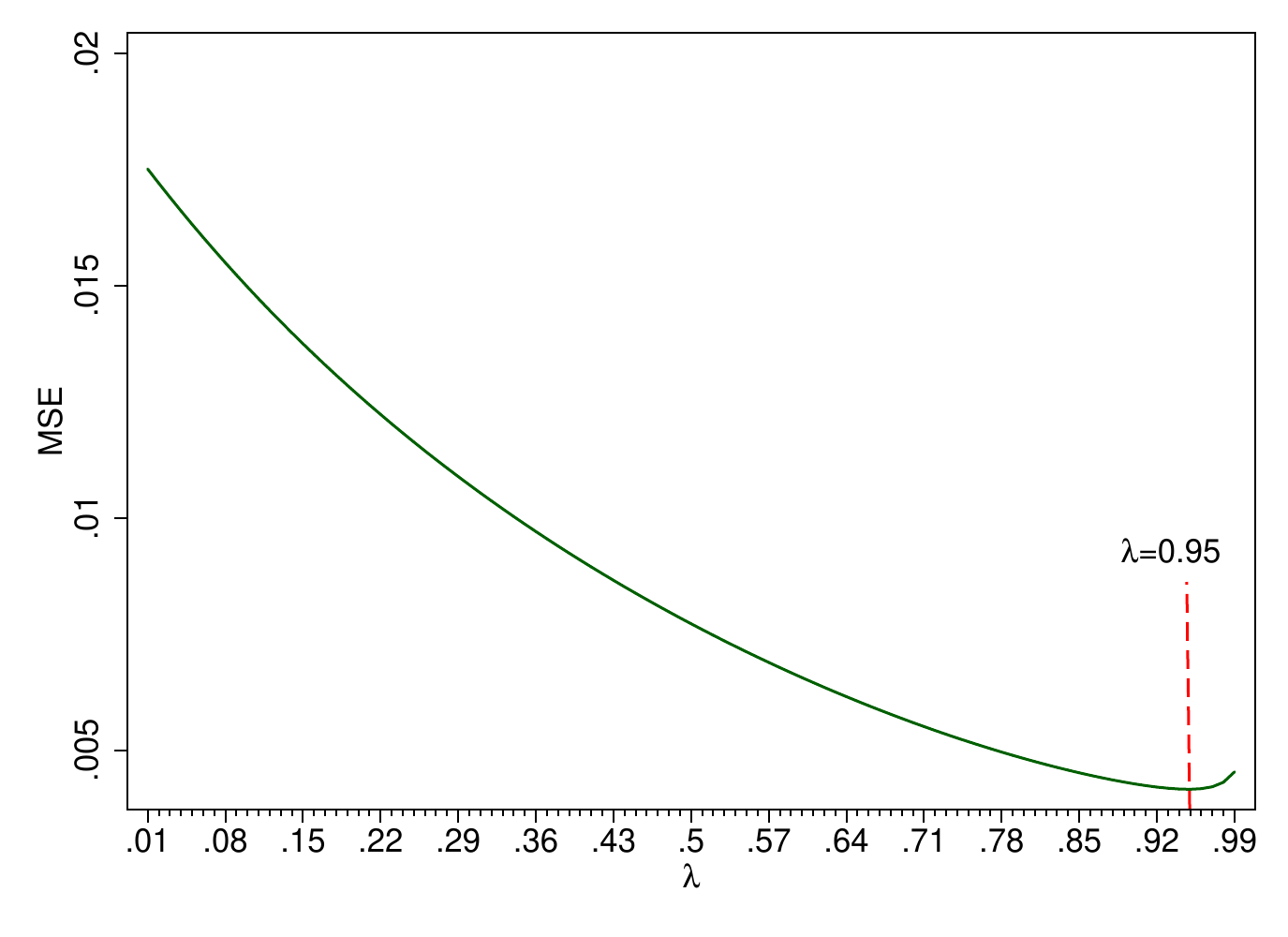}}
\end{figure}

Figures \ref{fig:One-week-forecast} and \ref{fig:Two-week-forecast}
display the results for 5 and 10 days as forecasting horizons finding
optimal decays equal to $\lambda_{5}^{*}=0.92$ and $\lambda_{10}^{*}=0.95$.
It confirms that the optimal decay parameter is a function of the
forecasting horizon. Besides, the decrease in $\lambda_{T}^{*}$ when
$T$ decline, implies a diminution in the persistence in favor of
short-memory. 

On a second approach, in addition to the full-sample optimal ( $\lambda_{T}^{*}$)
we also evaluated the use of the one-time lagged optimal decay parameter;
i.e, $\lambda_{(t-1)_{T}}^{*}$; to compute the covariance-matrix
at time $t$ taking into account the realizations at ($t-1$). For
illustrative purposes, Fig. \ref{fig:Optimal-weekly-1} plots the
daily optimal decay parameter for the monthly forecast horizon. Table
\ref{tab:MSE-using-} shows, for the three selected forecasting horizons,
the MSE under the two used approaches; i.e., the full-sample optimal
smoothing parameter and the time-varying one that minimizes the SE
at the immediately previous time. We can observe that there is an
improvement in the MSE loss function if we consider $\lambda_{(t-1)_{T}}^{*}$
instead of the constant-valued $\lambda_{T}^{*}$. To analyze if the
results are statistically significant or not, the $t$-statistics
of the pairwise \citeauthor{diebold1995comparing} are reported in
the table (positive sign indicates a better accuracy of the in-sample
time-varying optimal decay parameter), finding a superior forecasting
performance when we use $\lambda_{(t-1)_{T}}^{*}$ in place of $\lambda_{T}^{*}$. 

\begin{table}
\begin{centering}
\textbf{\caption{\textbf{MSE and $t-$Statistics of Diebold-Mariano (DM) joint test
for $\lambda_{t}=\lambda_{T}^{*}$ and $\lambda_{t}=\lambda_{(t-1)_{T}}^{*}$.\label{tab:MSE-using-}}}
\medskip{}
}
\par\end{centering}
\begin{centering}
\begin{tabular}{ccccccccc}
\hline 
 &
\multicolumn{2}{c}{\textbf{$\boldsymbol{T}$=5}} &
 &
\multicolumn{2}{c}{\textbf{$\boldsymbol{T}$=10}} &
 &
\multicolumn{2}{c}{\textbf{$\boldsymbol{T}$=21}}\tabularnewline
\cline{2-9} \cline{3-9} \cline{4-9} \cline{5-9} \cline{6-9} \cline{7-9} \cline{8-9} \cline{9-9} 
 &
\textbf{\scriptsize{}$\boldsymbol{\lambda_{t}=0.92}$} &
\textbf{\scriptsize{}$\boldsymbol{\lambda_{t}=\lambda_{(t-1)_{T}}^{*}}$} &
 &
\textbf{\scriptsize{}$\boldsymbol{\lambda_{t}=0.95}$} &
\textbf{\scriptsize{}$\boldsymbol{\lambda_{t}=\lambda_{(t-1)_{T}}^{*}}$} &
 &
\textbf{\scriptsize{}$\boldsymbol{\lambda_{t}=0.98}$} &
\textbf{\scriptsize{}$\boldsymbol{\lambda_{t}=\lambda_{(t-1)_{T}}^{*}}$}\tabularnewline
\hline 
\textbf{MSE} &
0.001355 &
0.001160 &
 &
.004159 &
.003056 &
 &
.01486 &
.01136\tabularnewline
\textbf{DM} &
\multicolumn{2}{c}{3.214{*}{*} } &
 &
\multicolumn{2}{c}{5.895{*}} &
 &
\multicolumn{2}{c}{5.174{*}}\tabularnewline
\end{tabular}
\par\end{centering}
\centering{}{\footnotesize{}\vspace{.5em} {*} Significant at $p$<0.001,
{*}{*} Significant at $p$<0.005}{\footnotesize\par}
\end{table}

\section{Summary}

In this paper, the EWMA volatility model is addressed by studying
the optimal decay parameter under different time horizons: 1 week,
2 weeks, and 1 year. Using the historical price database of the top
22 blue-chip companies in the US stock market in the period 2000-2020,
we evaluate the forecasting performance on that time-span through
the squared error loss function of the variance-covariance matrix. 

If we look for a full-sample optimal smoothing parameter, the results
are in agreement with the original RiskMetrics suggestion for the
1-month forecasting, finding an optimal value equal to 0.98 (RiskMetrics
suggest 0.97, which in our analysis ranked second with a very minimal
difference in the MSE). However, for daily forecast, our findings
report that the recent lags should receive more weight ($\lambda_{1}^{*}=0.89$)
than the RiskMetrics recommendation ($\lambda_{1}^{*}=0.94$); Nevertheless,
despite the lower MSE of the former, there are no statistical differences
in terms of forecasting accuracy. We also obtain the optimal full-sample
decay parameter considering 1-week and 2-week as forecasting horizons:
$\lambda_{5}^{*}=0.92$ and $\lambda_{10}^{*}=0.95$, respectively.

Unlike other research outputs where the estimation of $\lambda_{T}^{*}$
relies on the single-variance analysis of one or more time series,
here the full variance-covariance matrix is estimated considering
22 time-series (i.e., 253 entries per day) and 21 years of data.

Moreover, we also tested the use of the optimal value of $\lambda$
at time $t-1$ (namely $\lambda_{t-1}^{*}$) to predict the covariance
by EWMA for the time $t.$ We found an improvement in the MSE using
this way instead of a fixed $\lambda$, with the exception of the
1-day forecasting. On one hand, this improvement in the MSE attempts
to the simplicity of the EWMA approach because at every time we need
to run the EWMA model under a different set of $\lambda$ values and
select the optimal in each computation. On the other hand, this approach
takes much less time than other multivariate models (for example M-GARCH)
and could be developed without high computational skills (for instance,
in an Excel spreadsheet).

\section{Acknowledgments}

This work was supported from Operational Programme Research, Development
and Education - Project ``Postdoc2MUNI'' (No. \href{https://www.muni.cz/en/research/projects/56527}{CZ.02.2.69/0.0/0.0/18\_053/0016952}).

\bibliographystyle{plainnat}
\bibliography{11_Users_axelaraneda_Desktop_Research_Portfolio_port}

\end{document}